\documentclass[fleqn,usenatbib,useAMS]{mnras}

\usepackage{newtxtext,newtxmath}
\usepackage[T1]{fontenc}
\DeclareRobustCommand{\VAN}[3]{#2}
\let\VANthebibliography\thebibliography
\def\thebibliography{\DeclareRobustCommand{\VAN}[3]{##3}\VANthebibliography}
\usepackage{pdflscape}	
\usepackage{ae,aecompl}

\usepackage{graphicx} 

\newcommand{\clfd}{\mbox{\textsc{clfd}}}
\newcommand{\dspsr}{\mbox{\textsc{dspsr}}}
\newcommand{\presto}{\mbox{\textsc{presto}}}
\newcommand{\tempotwo}{\mbox{\textsc{tempo2}}}
\newcommand{\psrchive}{\mbox{\textsc{psrchive}}}
\newcommand{\pulsarX}{\mbox{\textsc{pulsarX}}}
\newcommand{\peasoup}{\mbox{\textsc{peasoup}}}
\newcommand{\pics}{\mbox{\textsc{pics}}}
\newcommand{\iqrm}{\mbox{\textsc{iqrm}}}
\newcommand{\dedisp}{\mbox{\textsc{dedisp}}}
\newcommand{\mosaic}{\mbox{\textsc{mosaic}}}
\newcommand{\transientX}{\mbox{\textsc{transientX}}}
\newcommand{\heimdall}{\mbox{\textsc{heimdall}}}
\newcommand{\riptide}{\mbox{\textsc{riptide}}}
\newcommand{\multiTRAPUM}{\mbox{\textsc{multiTRAPUM}}}
\newcommand{\psrfoldfil}{\mbox{\texttt{psrfold\_fil}}}
\newcommand{\aplpy}{\mbox{\textsc{aplpy}}}

\newcommand{\accelunits}{\,m\,s$^{-2}$}
\newcommand{\dmunits}{\,pc\,cm$^{-3}$}

\newcommand{\h}{$^{\rm h}$} 
\newcommand{\m}{$^{\rm m}$}

\newcommand{\thepulsar}{PSR\,J0058$-$7218}
\newcommand{\thepulsarnopsr}{J0058$-$7218}
\newcommand{\pulsarposition}{\mbox{RA(J2000) $=$ 00\h58\m16\fs85} \mbox{Dec(J2000) $=$ $-$72\textdegree{}18\arcmin05\farcs60} \mbox{$\pm$ 0\farcs6}}
\newcommand{\PWNposition}{\mbox{RA(J2000) $=$ 00\h58\m16\fs824} \mbox{$\pm$ 0\farcs1} \mbox{Dec(J2000) $=$ $-$72\textdegree{}18\arcmin05\farcs32} \mbox{$\pm$ 0\farcs2}}
\newcommand{\PWNpositionnoerrors}{\mbox{RA(J2000) $=$ 00\h58\m16\fs824} \mbox{Dec(J2000) $=$ $-$72\textdegree{}18\arcmin05\farcs32}}

\title[TRAPUM upper limits for \thepulsar]{TRAPUM upper limits on pulsed radio emission for SMC X-ray pulsar \thepulsarnopsr}

\author[E. Carli et al.]{\parbox{\textwidth}{
E. Carli,$^{1}$\thanks{E-mail: \href{mailto:emma.carli@postgrad.manchester.ac.uk}{emma.carli@postgrad.manchester.ac.uk}}
L. Levin,$^{1}$
B. W. Stappers,$^{1}$
E. D. Barr,$^{2}$
R. P. Breton,$^{1}$
S. Buchner,$^{3}$
M. Burgay,$^{4}$
M. Kramer,$^{2}$
P. V. Padmanabh,$^{2,5}$
A. Possenti,$^{4}$
V. Venkatraman Krishnan,$^{2}$
J. Behrend,$^{2}$
D.~J.~Champion,$^{2}$
W. Chen,$^{2}$
Y. P. Men$^{2}$
}
\\ \\ \\
$^{1}$Jodrell Bank Centre for Astrophysics, Department of Physics and Astronomy, The University of Manchester, Manchester M13 9PL, UK \\
$^{2}$ Max-Planck-Institut f\"{u}r Radioastronomie, Auf dem H\"{u}gel 69, D-53121 Bonn, Germany \\
$^{3}$ South African Radio Astronomy Observatory (SARAO), 2 Fir Street, Black River Park, Observatory, Cape Town, 7925 \\
$^{4}$INAF-Osservatorio Astronomico di Cagliari, via della Scienza 5, 09047, Selargius, Italy \\
$^{5}$Max-Planck-Institut f\"{u}r Gravitationsphysik (Albert-Einstein-Institut), D-30167 Hannover, Germany\\
}

\date{Accepted XXX. Received YYY; in original form ZZZ}
\pubyear{2022}

\begin{document}
\label{firstpage}
\pagerange{\pageref{firstpage}--\pageref{lastpage}}
\maketitle

\begin{abstract}
The TRAPUM collaboration has used the MeerKAT telescope to conduct a search for pulsed radio emission from the young Small Magellanic Cloud pulsar \thepulsarnopsr{} located in the supernova remnant IKT\,16, following its discovery in X-rays with \textit{XMM-Newton}. We report no significant detection of dispersed, pulsed radio emission from this source in three 2-hour L-band observations using the core dishes of MeerKAT, setting an upper limit of 7.0\,$\upmu$Jy on its mean flux density at 1284\,MHz. This is  nearly 7 times deeper than previous radio searches for this pulsar in Parkes L-band observations. This suggests  that the radio emission of \thepulsar{} is not beamed towards Earth or that \thepulsar{} is similar to a handful of Pulsar Wind Nebulae systems that have a very low radio efficiency, such as PSR\,B0540$-$6919, the Large Magellanic Cloud Crab pulsar analogue. We have also searched for bright, dispersed, single radio pulses and found no candidates above a fluence of 93\,mJy\,ms at 1284\,MHz.
\end{abstract}


\section{Introduction}
\label{introduction}
The Magellanic Clouds are the only galaxies outside our own in which radio pulsars have been discovered to date. They are nearby galaxies that are unobstructed by the galactic plane, therefore they are a good target for extragalactic pulsar searches. Indeed, the Small Magellanic Cloud (SMC) is just 60\,kpc away \citep{Karachentsev2004}, and the expected Milky Way Dispersion Measure (DM) contribution in its direction is low: about $30$\dmunits{}  according to the YMW2016 electron density model \citep{YMW2016}, and $42$\dmunits{} according to the NE2001 model \citep{NE2001}. 
There was a recent episode of stellar formation in the SMC about 40\,Myr ago \citep{Harris2004}, which led to an abundance of young systems per unit mass compared to the Milky Way. In particular, the SMC hosts many supernova remnants (18 confirmed SNRs in \citealt{Maggi2019}), which can contain young neutron stars. Due to their rarity in the Milky Way's older population\footnote{In the Australia Telescope National Facility (ATNF) \href{http://www.atnf.csiro.au/research/pulsar/psrcat}{pulsar catalogue} \citep{ATNF}, only about $\simeq$180 pulsars out of $\simeq$2500 with a catalogued characteristic age are younger than 120\,kyr.}, the discovery of young pulsars is scientifically interesting. Among them, magnetars (which have the highest known stellar magnetic fields) are prized for a variety of fundamental physics and astrophysics investigations \citep{Esposito2021}, including as one of the progenitors of Fast Radio Bursts  \citep{Bochenek2020, Caleb2021}. Each Magellanic Cloud hosts one known magnetar \citep{Helfand1979,Lamb2002}. Additionally, extragalactic pulsars are sought after for insights on the impact of different galactic properties on neutron star formation, such as the impact of metallicity and star formation history on stellar mass and supernova physics \citep{Heger2003,Titus2020}. 

There are seven published radio pulsars in the SMC and all have been discovered with the Parkes 64-m single dish radio telescope in Australia, by  \cite{McConnell1991} (one pulsar), \cite{Crawford2001} (1), \cite{Manchester2006} (3), and  \cite{Titus2019} (2). Their median DM is around 115\dmunits{}. 
The radio emission from these pulsars is powered by their rapid rotation. This can also power hard X-ray emission when the pulsar is young and most rapidly rotating. The X-ray emission has a characteristic hard power-law spectrum \citep[and references therein]{Becker1997}. 
Only two young\footnote{All other published non-magnetar pulsars in the SMC are over 1\,Myr in characteristic age \citep{ATNF}.} pulsars have been discovered in the SMC, both in the X-ray band only:  CXOU\,J010043.1$-$721134  \cite[a magnetar]{Lamb2002}, and most recently, \thepulsar{} \citep{Maitra2021}. In this paper, we describe the follow-up radio observations of the latter with MeerKAT.

\thepulsar{} is situated in the supernova remnant (SNR) IKT\,16 \citep{Inoue1983}, also known as DEM\,S\,103 \citep{Davies1976}. 
The remnant was first identified as an $\rm{H_\upalpha}$ emission nebula LHA\,115-N\,66 in \cite{Henize1956}. The first radio observations of the nebula were made by \cite{Mills1971}, and it was tentatively suggested to be a nonthermal radio source by \cite{Clarke1976}. It was catalogued as 1E0057.6$-$7228 with the first X-ray detection of the system by \cite{Seward1981}. \cite{Mills1982} conducted further radio observations which led to the proposition that DEM\,S\,103 was a tentative SNR candidate. It was officially declared a candidate SNR based on soft X-ray observations by \cite{Inoue1983}, and subsequently named IKT\,16. A radio and $\rm{H_\upalpha}$ study by \cite{Mathewson1984} confirmed that IKT\,16 is a SNR, which they found is in the Sedov evolutionary phase. 
Further insights as to the nature of the system were made possible by the launch of the \textit{XMM-Newton} and \textit{Chandra} X-ray observatories. \cite{VanDerHeyden2003} identified a hard X-ray source within the SNR in a single \textit{XMM} exposure, but it was too faint to be further characterised. After 8 additional \textit{XMM-Newton} observations,  \cite{Owen2011}  identified the hard X-ray source as a tentative Pulsar Wind Nebula (PWN), within a larger, extended radio source from ATCA and MOST radio images (inside the SNR, see \autoref{fig:ds9}). As the PWN is offset from the centre of the SNR, they suggested the existence of a pulsar in motion, with a transverse kick velocity of 580\,km\,s$^{-1}$. Using adiabatic Sedov modelling, they estimated the age of the SNR to be 14.7\,kyr. The putative pulsar emission could not be resolved from the PWN at the time.

\cite{Maitra2015} confirmed the existence of the PWN with \textit{Chandra} observations, resolving a putative pulsar at its centre in hard X-rays.  IKT\,16 was thus defined as a `composite SNR', where there is non-thermal emission from both the SNR shell in radio and the hard X-ray point source within the PWN. 
The latter's measured X-ray spectral index implied a young, rotation-powered pulsar with non-thermal radiation from the pulsar magnetosphere \citep[e.g.][]{Becker1997, Wang2004}. 
Using a similar method as \cite{Kargaltsev2008} using the X-ray luminosities of the PWN and point source, \cite{Maitra2015} estimated the period of the putative pulsar to be of the order of 100\,ms with a period derivative  $\simeq10^{-13}$.
\cite{Maitra2015}   characterised the morphology of the system: the PWN is a 5.2 arcsecond elongated X-ray feature at \PWNposition{} (\textit{Chandra} position). The putative pulsar (hard X-ray point source) is consistent with the PWN centre,  at \pulsarposition{} (\textit{Chandra} position). The PWN is located within a larger radio feature, which \cite{Maitra2015} described as two lobes about the PWN each 20$\pm$5 arcseconds in extent (see \autoref{fig:ds9}), from a high resolution ATCA 2.1\,GHz image. They concluded that the PWN is in expansion in the cold SNR ejecta. No radio point source was detected near the X-ray point source.

The first dedicated search for a radio pulsar in IKT\,16 was conducted by \cite{Titus2019} with the Parkes radio telescope, in two four-hour observations with a beam of the Parkes Multi-Beam receiver  \citep{Staveley-Smith1996} placed 1.4 arcminutes from the position of the putative pulsar (see \autoref{fig:ds9}). The observations had a 400 MHz bandwidth centred on 1382\,MHz (Berkeley–Parkes–Swinburne data recorder: \citealt{McMahon2011,Keith2010}). They carried out a Fourier domain search with \presto{} \citep{PRESTO, ransom2002} for DMs up to 660\dmunits{} and a fast-folding search with \riptide{} for DMs up to 400\dmunits{} \citep{riptide}. \thepulsar{} was not detected down to S/N$=$8, corresponding to a 1400\,MHz flux density limit of 36\,$\upmu$Jy. This limit takes into account the sensitivity loss from the putative pulsar being off-centre in the beam and assumes a 5 per cent duty cycle. With a 7\,K sky temperature contribution included \citep{skytemperature}, the flux density limit is 47\,$\upmu$Jy.

The 21.7\,ms pulsar \thepulsarnopsr{} was finally discovered in X-rays by \cite{Maitra2021} in 2020 \textit{XMM-Newton} EPIC camera observations. They measured a hard X-ray spectral index of 1.4$\pm 0.1$ and a single peak pulse shape.  They determined an ephemeris using the pulsar position from \cite{Maitra2015}. The pulsar has a high period derivative of $\simeq 3 \times 10^{-14}$, giving a young characteristic age of about 12\,kyr (in line with the SNR Sedov age from \citealt{Owen2011}), and a high  spin-down luminosity $\dot{E}=1.1 \times 10^{38}$\,erg\,s$^{-1}$. The inferred surface magnetic field of $B_{\text{surf}} =8 \times 10^{11}$\,G excludes that \thepulsar{} is a magnetar \citep{Esposito2021}. \cite{Maitra2021} conducted multiwavelength searches of \thepulsar{}: they reported no gamma-ray pulsations in the most recent \textit{Fermi} LAT data set, and no radio pulsations in a reprocessing of the \cite{Titus2019} Parkes data. They folded the data at the predicted topocentric period of the pulsar using \dspsr{} and searched the folded data  for the highest S/N profile over DMs 0-1000\dmunits{} with \psrchive{}'s \texttt{pdmp} tool \citep{DSPSR,psrchive_psrfits}. They quote a better flux density limit than \cite{Titus2019}, 15\,$\upmu$Jy at 1400\,MHz, as they use a lower signal-to-noise ratio (S/N) limit of 5 in their folding search, and do not take into account the sensitivity loss from the pulsar being off-centre in the Parkes beam (M. Pilia, private communication). Adding this sensitivity loss, a sky temperature contribution of 7\,K, and a S/N cut of 7 to compare with our own searches, this limit becomes 41\,$\upmu$Jy at 1400\,MHz. They also did not find any single pulses in this dataset with \heimdall{} over DMs 0-1000\dmunits{} \citep{Barsdell2012}. Using the survey parameters from \cite{Titus2019}, and taking into account the sensitivity loss from the pulsar being off-centre in the Parkes beam, a sky temperature contribution of 7\,K, and a S/N cut of 8 to compare with our own searches, we find their single pulse search has a limiting fluence of 821\,mJy\,ms\footnote{In \cite{Maitra2021}, a fluence limit of 160\,mJy\,ms down to S/N$=$5 is quoted. The beam correction and sky temperature are again not taken into account. Additionally, the correction factor accounting for data acquisition system imperfections is not used (M. Pilia, private communication).} for a 1\,ms pulse width at 1400\,MHz (with a 1\,ms integration time).

TRAPUM (TRAnsients and PUlsars with MeerKAT) is a Large Survey Project of the MeerKAT telescope (\href{http://trapum.org/}{trapum.org}, \citealt{Stappers2016}). One of TRAPUM's science goals is to find new extragalactic pulsars. Thus, the collaboration is currently conducting a survey of the SMC with the MeerKAT telescope. \thepulsar{} and its Pulsar Wind Nebula are located within several of the survey's pointings. The X-ray pulsar discovery in \cite{Maitra2021} was published during our survey, after one observation of the PWN was made.

\section{Observations and data reduction}
\label{Obs_and_processing}

\begin{figure}
\centering
\includegraphics[width=\columnwidth]{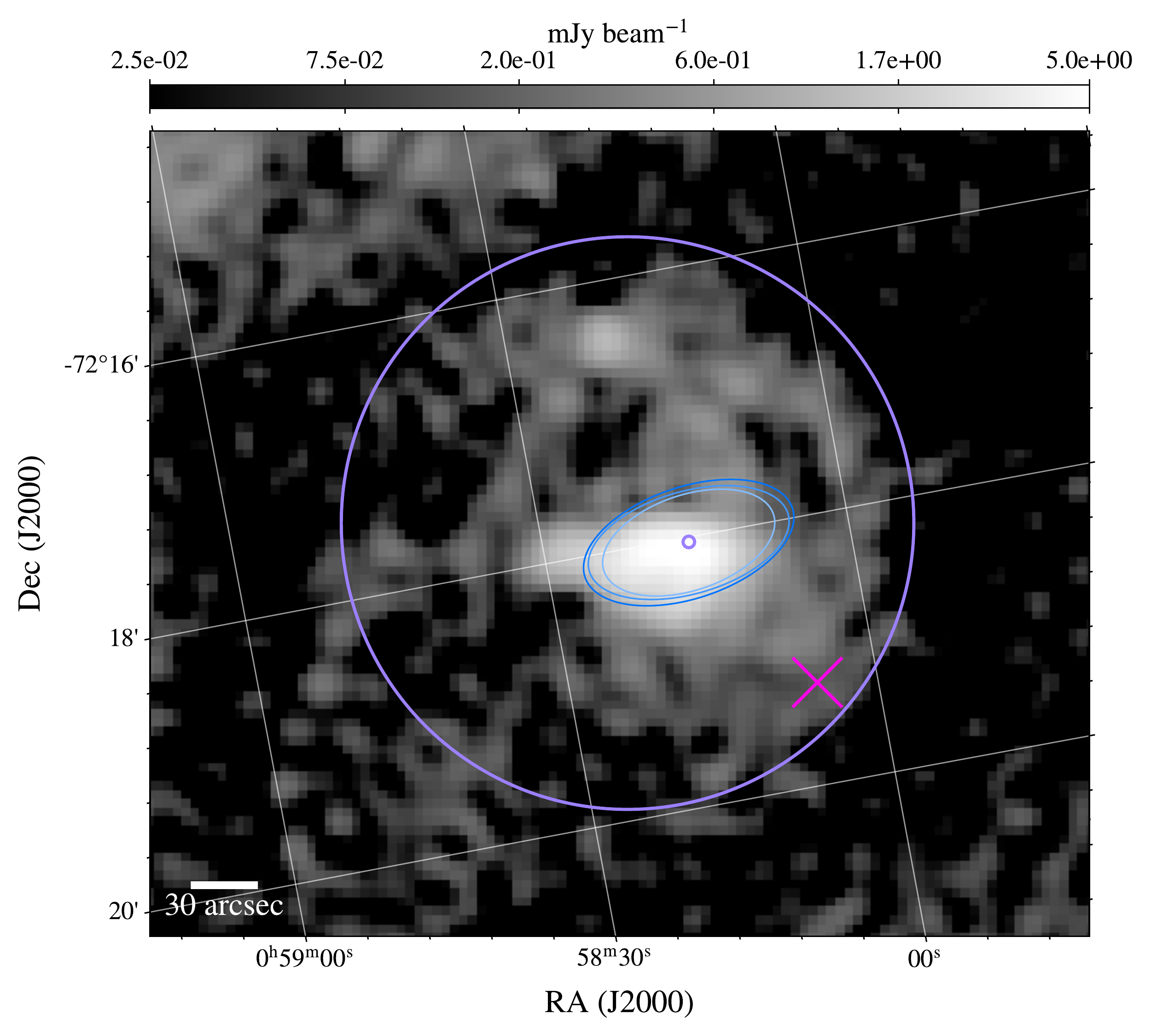}
\caption[caption:ds9]{Regions of interest are overlaid on a 1320\,MHz ASKAP image of the SMC \citep{Joseph2019}. The morphology of the composite SNR is delineated by purple circles. The largest purple circle delimits the SNR shell as catalogued by \cite{Maggi2019}. It contains a comparatively bright feature around the 5.2\,arcsecond diameter PWN (smallest purple circle, \citealt{Maitra2015}). Our observed MeerKAT beams are shown at 50 per cent sensitivity as blue ellipses, simulated with \mosaic{} \citep{Chen2021}. The innermost beam is OBS1 and the outermost beam is OBS3 (cf. \autoref{tab:observations}). They are centred on the PWN position. The centre of the Parkes telescope beam from the \cite{Titus2019} survey is shown as a pink cross. The 14\,arcminute Parkes half-power beam width lies outside the figure. This figure was generated with the \href{http://aplpy.github.io/}{\aplpy{}} Python package.
}
\label{fig:ds9}
\end{figure}

\begin{table*}
	\centering
	\caption{Characteristics of the MeerKAT coherent beam observations of \thepulsar{}. For each observation, one coherent beam is placed at the centre of the PWN as defined by \protect\cite{Maitra2015}, \PWNpositionnoerrors{}. The coherent beam is placed within the MeerKAT primary incoherent beam, which pointing location  is given in Right Ascension and Declination (J2000). The coherent beam major axes are given in arcseconds in the Right Ascension and Declination directions, with the ellipse position angle.}
	\label{tab:observations}

\resizebox{\textwidth}{!}{
\begin{tabular}{cccccc}
\hline
\textbf{Observation Number and Date}                        & \multicolumn{1}{l}{\textbf{Primary incoherent beam location}}                & \textbf{Number of dishes} & \multicolumn{1}{l}{\textbf{Coherent beam size (50 per cent sensitivity)}}                & \textbf{Sampling time} & \textbf{Processing}                                                               \\ \hline
\begin{tabular}[c]{@{}c@{}}OBS1\\ 14 Jan 2021\end{tabular} & \begin{tabular}[c]{@{}c@{}}0\h55\m14\fs6910\\ -72\textdegree{}26\arcmin06\farcs899\end{tabular} & 40                        & \begin{tabular}[c]{@{}c@{}}40\farcs0, 21\farcs3\\ 9\fdg2 \end{tabular} & 76\,$\upmu$s           & \begin{tabular}[c]{@{}c@{}}Pulsar search\\ Fold on downsampled data\end{tabular}   \\ \hline
\begin{tabular}[c]{@{}c@{}}OBS2\\ 24 Apr 2021\end{tabular}   & \begin{tabular}[c]{@{}c@{}}0\h55\m14\fs6910\\ -72\textdegree{}26\arcmin06\farcs899\end{tabular} & 44                        & \begin{tabular}[c]{@{}c@{}}46\farcs2, 22\farcs9\\ 6\fdg6 \end{tabular} & 153\,$\upmu$s          & \begin{tabular}[c]{@{}c@{}}Pulsar search\\ Fold\\ Single pulse search\end{tabular} \\ \hline
\begin{tabular}[c]{@{}c@{}}OBS3\\ 27 Oct 2021\end{tabular} & \begin{tabular}[c]{@{}c@{}}0\h57\m21\fs2370\\ -71\textdegree{}52\arcmin35\farcs195\end{tabular} & 44                        & \begin{tabular}[c]{@{}c@{}}48\farcs9, 24\farcs9\\ 9\fdg0 \end{tabular} & 153\,$\upmu$s          & \begin{tabular}[c]{@{}c@{}}Pulsar search\\ Fold\\ Single pulse search\end{tabular} \\ \hline
\end{tabular}%
}

\end{table*}

We placed one coherent beam at the centre of the PWN\footnote{\cite{Maitra2015} showed that the pulsar position is consistent with the centre of the PWN (see \autoref{introduction}).
} (as defined by \cite{Maitra2015}: \PWNpositionnoerrors{}),  in three multi-beam observations as part of our SMC survey. The beams were formed using the core dishes of MeerKAT (40 or 44 dishes out of 64) with a maximum baseline length of 1\,km. This reduction from the full array's 8\,km maximum baseline strikes a good balance between beam area and sensitivity \citep{Chen2021}. We carried out the three 2-hour observations at L-band (856-1712\,MHz, 856\,MHz bandwidth), with 2048 frequency channels, and a beam area large enough to cover the \textit{Chandra} hard X-ray pulsar position error: \pulsarposition{} \citep{Maitra2015}.  The characteristics of the individual observations are given in \autoref{tab:observations} and \autoref{fig:ds9}. 

We have run a pulsar search on all three observations of the PWN, as part of the standard processing used in the TRAPUM SMC survey. Radio Frequency Interference (RFI) cleaning was performed with a combination of \iqrm{} \citep{IQRM}, \presto{}'s \texttt{rfifind}, and  \pulsarX{}'s \texttt{filtool}\footnote{\href{https://github.com/ypmen/PulsarX}{https://github.com/ypmen/PulsarX} by Yunpeng Men}. As this was part of a multi-beam search we also used a multi-beam RFI filter \multiTRAPUM{}\footnote{\href{https://github.com/mcbernadich/multiTRAPUM}{https://github.com/mcbernadich/multiTRAPUM} by Miquel Colom i Bernadich}. The latter is a wrapper of \texttt{rfifind} which removes signals detected in several beams with sufficient spatial separation. The de-dispersion plan was computed with \presto{}'s \texttt{DDplan.py} script. We used the \dedisp{} library \citep{levin2012,Barsdell2012} for dedispersion over DMs from 50 to 350\dmunits{} as part of the \peasoup{}\footnote{\href{https://github.com/ewanbarr/peasoup}{https://github.com/ewanbarr/peasoup} by Ewan Barr} suite, a GPU-based time-domain linear acceleration  pulsar searching software. The DM range was chosen based on the median DM of the published pulsars in the SMC, around 115\dmunits{}. The PWN and SNR material could add a contribution to the DM. Using \peasoup{}, the 2-hour long data were searched with no acceleration, while 20-minute segments were searched with accelerations up to $|50|$\accelunits{}, to allow for binary systems\footnote{We already know from \cite{Maitra2021} and \cite{Ho2022} that the system is not significantly accelerated, but acceleration searching is part of standard TRAPUM SMC Survey processing, and OBS1 was processed before the discovery of \thepulsar{} was published in \cite{Maitra2021}.} (assuming constant acceleration). We chose an acceleration tolerance parameter of 10 per cent \citep{levin2012}. This means the acceleration broadening from one acceleration step to the next cannot exceed 10 per cent of the combined pulse smearing due to sampling time, intra-channel frequency dispersion, and de-dispersion step size. The 2-hour timeseries were zero-padded to the next nearest power of 2, $2^{26}$, resulting in a better resolved Fourier spectrum. Candidates were harmonically summed by \peasoup{} up to the eighth harmonic \citep{Taylor1969}. The resulting candidates were then sifted, i.e. clustered in DM, period, acceleration and harmonics. A first round of sifting was performed directly by \peasoup{}, which returned on average around 10\,000 candidates in the beam for one observation. It was followed by multi-beam spatial sifting\footnote{\href{https://github.com/prajwalvp/candidate_filter}{https://github.com/prajwalvp/candidate\_filter} by Lars K\"{u}nkel}, which uses the expected spatial relationship of real sources to identify RFI, after which about 3000 candidates remained. We decided to only keep candidates with periods  longer than 8 time samples (1.216\,ms) for OBS2 and OBS3, which had a sampling time of 153\,$\upmu$s (cf. \autoref{tab:observations}). For OBS1, that had a 76\,$\upmu$s sampling time, the shortest period folded was the \peasoup{} default minimum period searched of 0.91\,ms. The longest period searched by \peasoup{} is 10 seconds. We folded about 1000  \peasoup{} candidates in these period ranges with a spectral S/N above 9 using \pulsarX{}'s \psrfoldfil{}. \psrfoldfil{} cleans the full resolution data before folding and applies \clfd{} on the folded data\footnote{\href{https://github.com/v-morello/clfd}{https://github.com/v-morello/clfd}  by Vincent Morello}. The folds were searched for the highest S/N profile over a so-called `natural' range around the candidate period, period derivative, DM, and acceleration. We then partially classified the folded candidates  with \pics{} \citep[a Pulsar Image-based Classification System based on Machine Learning]{PICS}. We used a minimum score of 10 per cent pulsar-like for  both the original and TRAPUM training sets \citep{VoragantiPadmanabh2021}, as well as a minimum folded S/N of 7. This returned a handful of candidates in the beam for each observation, all of which were consistent with RFI or noise. The original \peasoup{} candidates files, where the spectral S/N was limited to above 8, were inspected in case the pulsar candidate had a lower spectral S/N than our folding cut-off of 9, or was lost in sifting, but no periodic signals were found at \thepulsar{}'s period. More information about the TRAPUM search pipeline can be found in Chapter 3 of \cite{VoragantiPadmanabh2021}.

The discovery of \thepulsar{} was published in \cite{Maitra2021} after the full resolution data for OBS1 were deleted\footnote{As the raw data are too voluminous to keep, we stored a RFI-cleaned downsampled (512 channels each de-dispersed at a DM of 115\,\dmunits{} and 1.225\,ms sampling time) version of OBS1.}. We used the ephemeris provided\footnote{Even if the ephemeris did not extend to our observations, the pulsar search could detect the source within our survey pipeline limits.} to fold a downsampled OBS1 and the full resolution OBS2 and OBS3 with \dspsr{} (which calls on \tempotwo{} \citep{TEMPO2_overview} to predict the pulsar period). The folded data were cleaned with \presto{}'s \texttt{rfifind} frequency channel mask from the full resolution data, \clfd{}, and \psrchive{}. We then searched for the highest S/N profile in the folded data with \psrchive{}'s \texttt{pdmp} tool up to a DM of 500\dmunits{}, with a period range of $\pm$1\,$\upmu$s, which is three orders of magnitude greater than the period error determined by \cite{Maitra2021}. This should also cover any period derivative inaccuracies. We also ran folding searches up to a DM of 1000\dmunits{} with \pulsarX{}'s \psrfoldfil{}  on the downsampled data from OBS1 and the full resolution OBS2 and OBS3 data. We applied a \texttt{rfifind} frequency channel mask from the full resolution data. \psrfoldfil{} cleans the full resolution data before folding and applies \clfd{} on the folded data. The folds were searched for the highest S/N profile over a so-called `natural' range around the ephemeris-predicted topocentric spin frequency $\left(\pm\frac{1}{\text{T}_{\text{obs}}}\right)$ and spin-down rate $\left(\pm\frac{2}{\text{T}_{\text{obs}}^2}\right)$, where $\text{T}_{\text{obs}}$ is the observation length in seconds. This is 3 and 4 orders of magnitudes larger than the error bounds from \cite{Maitra2021} respectively. Both folding procedures were split into several searches, each centred on a DM, with a range of $\pm 5$\dmunits{} for \texttt{pdmp}, and a DM range so that the maximum dispersion delay is 1 pulse period over the whole observation for \psrfoldfil{}. The DMs were spaced by 10\dmunits{} over the entire DM range, and returned no significant pulsations at the appropriate parameters.

A new ephemeris for \thepulsar{} was published in \cite{Ho2022} from \textit{NICER} X-ray observations \citep{NICER}. This ephemeris has smaller error ranges on its parameters as it covers an 8-month X-ray timing period, while the ephemeris from \cite{Maitra2021} used data from 2020 \textit{XMM-Newton} EPIC camera observations that only spanned 1.4 days. This consolidated the proposed age and $\dot{E}$ of \thepulsar{}. \cite{Ho2022} noted that this is the fourth highest $\dot{E}$ known.  The new ephemeris also contains a second period derivative.  We repeated the same folding search procedure as before with this new ephemeris, except for the  \texttt{rfifind} frequency channel mask which was not applied (\pulsarX{}'s \texttt{filtool} was run directly on the data instead) and the period derivative plane was not searched (only DM and period). We also used higher resolution folded data. Again, there were no significant pulsations at the appropriate parameters.

Young pulsars like \thepulsar{} can emit giant narrow pulses \citep[e.g. the LMC Crab pulsar twin,][]{Johnston2003}. If \thepulsar{} is too faint to be detected by MeerKAT in 2-hour observations, single pulses that are orders of magnitude stronger than average may still be detected. We thus searched for single pulses with \transientX{}\footnote{\href{https://github.com/ypmen/TransientX}{https://github.com/ypmen/TransientX} by Yunpeng Men} on the full resolution data from OBS2 and OBS3, cleaned by \texttt{rfifind} and \pulsarX{}'s \texttt{filtool}. We used a 100\,ms maximum search width over DMs 0 to 5000\dmunits{}. No pulses were returned down to S/N$=$8. Similarly, we ran a \heimdall{} search on the same data, cleaned by \iqrm{} and \texttt{rfifind}.  We chose a maximum boxcar width of 157\,ms over DMs 25 to 5000\dmunits{}. We discarded any pulses with widths wider than 50 per cent of \thepulsar{}'s period, as giant pulses are typically narrow \citep[e.g.][]{Crawford2009}. We used \dspsr{} and \psrchive{} to extract single pulses, search in DM for the highest S/N profile and plot the candidates.  Inspection of all the candidates above S/N$=$8 indicated they were consistent with RFI or noise.

\section{Upper limits and discussion}

We calculate a radio L-band flux density upper limit for \thepulsar{} with the folds performed on the data from OBS2, in which the pointing location of MeerKAT's primary incoherent beam was closest to the pulsar and 44 dishes were used (see \autoref{tab:observations}). Owing to the separation between the coherent beam placed on \thepulsar{} and the centre of the primary beam, the gain was decreased to $\simeq$85 per cent of its full value (using a primary beam model based on \citealt{Asad2021}).  Taking this into account, the radiometer equation applied to pulsar observations \citep{handbook} 
yields a flux density upper limit of $S_{\text{1284\,MHz}} =$ 7.0\,$\upmu$Jy. We assume a 5 per cent pulsar duty cycle and a minimum folded S/N of 7 to be able to distinguish the signal by eye.  We use the MeerKAT system temperature ($\simeq$18\,K), a sky temperature of 7\,K \citep{skytemperature}, and a gain of $\simeq$1.925\,K Jy$^{-1}$ for the 44 core dishes \citep{Bailes2020} at the centre of the pointing (primary beam location). We assume that a correction factor accounting for any data acquisition system imperfections is not needed. Using the flux density from the ASKAP image shown in \autoref{fig:ds9}, we take the faint remnant's contribution to the system temperature as negligible. 
Assuming a power law radio spectral index of -1.60$\pm$0.54 \citep{Jankowski2018}, the radio flux density upper limits at 1400\,MHz are $S_{\text{1400\,MHz}} =$ 6.1$\pm$0.3\,$\upmu$Jy\footnote{\cite{Titus2020} calculated the TRAPUM search limiting flux density to be 12\,$\upmu$Jy at 1400\,MHz with different assumptions, notably a smaller bandwidth, higher S/N cut  and lower sensitivity (N. Titus, private communication).}.  Our 1400\,MHz flux density limit is $\simeq$6.7 times deeper than the 41\,$\upmu$Jy limit from \cite{Maitra2021}. 

Similarly, the limiting fluence for the single pulse search on OBS2 down to S/N$=$8 is $S_{\text{pulse,1284\,MHz}} =$ 93\,mJy\,ms, for a 1\,ms pulse width (and a 1\,ms integration time). This takes into account the primary beam sensitivity loss and the sky temperature contribution. Again, this greatly improves the previous upper limit from \cite{Maitra2021} of 821\,mJy\,ms, resulting from a transient search on \cite{Titus2019} Parkes data with a central frequency of 1382\,MHz.

If \thepulsar{}'s radio beam is sweeping across our line of sight, the pulsar could be too faint to be detected in 2 hours with the core dishes of MeerKAT. The upper limit on the flux density of the pulsar at 1400\,MHz translates to a radio pseudo-luminosity upper limit of $L_{\text{pseudo,1400\,MHz}}= S_{\text{1400\,MHz}} \times D^{2} = $ 22\,mJy\,kpc$^{2}$, assuming an approximate distance to the pulsar of $D=$ 60\,kpc \citep{Karachentsev2004}, and a power law radio spectral index of -1.6. According to the ATNF pulsar database, for pulsars with a catalogued pseudo-luminosity, approximately 75 per cent of all  pulsars and more than 50 per cent of young pulsars with a characteristic age lower than 120\,kyr have a lower pseudo-luminosity than this upper limit for \thepulsar{}  \citep{ATNF}.
Since we did not detect any giant single pulses either, \thepulsar{} was either not emitting giant pulses during our observations or its radio beam is not sweeping our line of sight.

\cite{Maitra2021} measured a pulsed, non-thermal\footnote{The pulsed X-rays are from the pulsar's emission beam, not from any heating processes from the pulsar surface or nebula.} X-ray luminosity for \thepulsar{} of $1.6(1)\times 10^{35}$\,erg\,s$^{-1}$ in the energy band 0.2-12\,keV, assuming a distance of 60\,kpc \citep{Karachentsev2004}. Furthermore, they determined the pulsar spin-down luminosity $\dot{E}=1.1 \times 10^{38}$ \,erg\,s$^{-1}$. Thus, they state that the X-ray radiation efficiency of the pulsar, $\eta_{\text{x}}=\frac{L_{\text{x}}}{\dot{E}}$ is of the order of $10^{-3}$. This fits well with the expected relationship between $L_{\text{x}}$ and $\dot{E}$ \citep{Shibata2016}. The ratio of \thepulsar{}'s X-ray luminosity to its radio luminosity upper limit, $\frac{L_{\text{x}}}{L_{\text{radio}}}$,  is of the order of $10^{5}$.
\cite{Maitra2021} and \cite{Ho2022} did not detect any statistically significant pulsed gamma-ray flux from \thepulsar{}. They could not provide a pulsed gamma-ray luminosity upper limit, so it is not possible to constrain the gamma-ray efficiency. 

Using a general formula for pulsar luminosity derived in \cite{handbook} and used in \cite{Szary2014} to broadly compare the radio emission efficiencies of pulsars, we find that the radio luminosity upper limit for \thepulsar{} is  $L_{\text{radio}} = 7.4  \times 10^{27} \times  L_{\text{pseudo,1400\,MHz}} = 1.6\pm0.1 \times 10^{29}$\,erg\,s$^{-1}$. Thus the radio radiation efficiency of the pulsar is lower than $\eta_{\text{radio}}=\frac{L_{\text{radio}}}{\dot{E}} = 1.5\pm0.1 \times 10^{-9}$. The ranges are due to the assumed radio spectral index,  -1.60$\pm$0.54 \citep{Jankowski2018}.  According to the ATNF pulsar database, only eight pulsars out of $\simeq$ 2000 with catalogued radio and spin-down luminosities have a lower radio radiation efficiency than this upper limit, although we note that it is speculative to use this general pulsar luminosity formula to compare individual pulsars, especially with uncertain distance estimates. They are all young\footnote{The largest characteristic age of the eight pulsars is 42\,kyr.} Milky Way pulsars with pulsed high-energy emission (gamma-ray and/or X-ray), high $\dot{E}$ ($> 10^{36}$ \,erg\,s$^{-1}$) and most have associated Pulsar Wind Nebulae and/or supernova remnants. Notably, the Crab pulsar is among these systems. It is the only pulsar out of the 8 with known giant pulses. Those detected in X-ray pulsations all have an X-ray efficiency $\eta_{\text{x}}$ that fits well with the expected relationship between $L_{\text{x}}$ and $\dot{E}$ \citep{Shibata2016}.  \thepulsar{} could thus be a similar system with a very low radio emission efficiency. \cite{Szary2014} show that radio efficiency generally increases with pulsar age, due to the constancy of their radio luminosity and decline in spin-down luminosity, which is a important insight into the radio emission mechanism.

\begin{figure}
\centering
\includegraphics[width=\columnwidth]{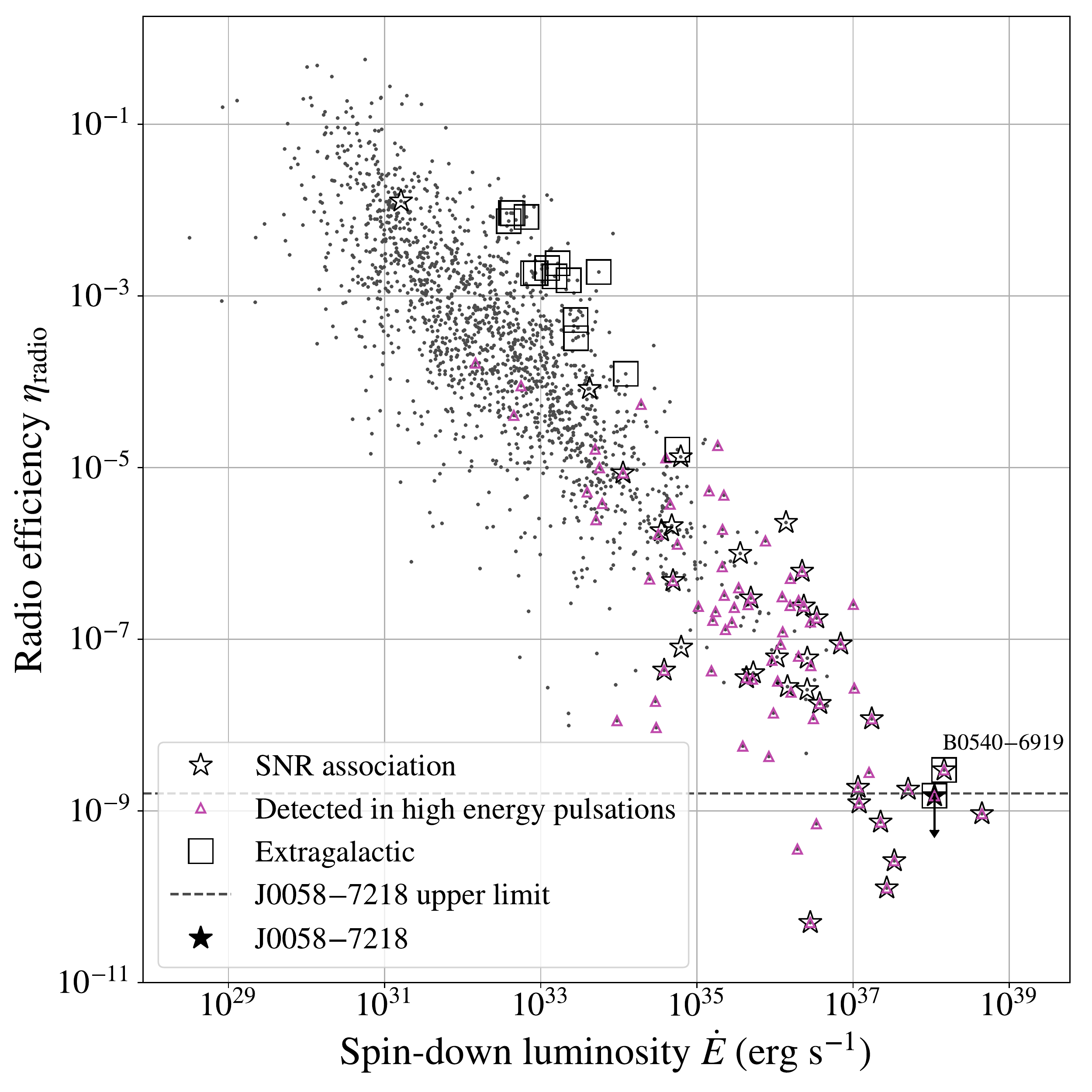}
\caption[caption:efficiency-Edot-plot]{ The $\eta_{\text{radio}}$-$\dot{E}$  relationship adapted from  \cite{Szary2014}. \thepulsar{} and LMC `Crab pulsar twin' PSR B0540$-$6919 are highlighted. As discussed in the text, it is speculative to use a general pulsar luminosity formula to compare individual pulsars' radio efficiencies.}
\label{fig:efficiency-Edot-plot}
\end{figure}

Indeed, when considering the $\eta_{\text{radio}}$-$\dot{E}$ relationship presented in \cite{Szary2014} with the latest version of the ATNF pulsar catalogue,  as presented in \autoref{fig:efficiency-Edot-plot}, \thepulsar{}'s upper limit on $\eta_{\text{radio}}$ (and lower values down to $\eta_{\text{radio}} \simeq 10^{-11}$) fits  within the large relationship scatter, among young pulsars with pulsed high-energy radiation and SNR association. This plot was created with pulsars from the ATNF database with a catalogued distance and 1400\,MHz flux density to calculate $L_{\text{radio}}$; and with a catalogued period derivative to calculate $\dot{E}= 4 \pi^{2} \times  10^{45} \times \dot{P} P^{-3}$. 
The resulting radio emission efficiencies $0 < \eta_{\text{radio}} < 1$ were plotted, removing anomalous values. Note that only so-called `normal' spin-powered pulsars are shown, excluding millisecond pulsars, pulsars in Globular Clusters, Rapidly Rotating Radio Transients (RRATs), pulsars in binary systems, and magnetars.  

As described in \cite{Maitra2021}, the fact that no thermal X-rays are detected from the pulsar, only non-thermal, could mean that the polar cap region (where thermal X-rays are emitted due to accelerated ions colliding with the surface) is not visible. Hence the beam of the radio emission could be missed \citep{Goldreich1969,Cheng1998}, and no pulses or giant pulses detected.
The radio beam could also be  intersected at a wide angle where the emission is faint. Indeed, young, short period pulsars can have wide radio beaming angles \citep[e.g.][]{Lyne1988}.
Since the radio, X-ray and gamma-ray beams can all have different orientations, the gamma-ray beam could also be missed \citep[e.g][]{Barnard2016,Cheng1999}. According to \cite{Smith2019}, \thepulsar{} is near the putative gamma-ray emission `death line' (with $\dot{E}=1.1 \times 10^{38}$\,erg\,s$^{-1}$ \citep{Maitra2021} and a distance of 60\,kpc), and thus could have no gamma-ray emission detectable at all. In the case this `death line' is an artefact of pulsar distance, the gamma-ray emission could be too faint to be detected with \textit{Fermi}.

\thepulsar{} thus joins a growing population of young pulsars with no radio detection. In the current ATNF pulsar catalogue, only half of the $\simeq$50 young pulsars with SNR association are detected in both high-energy and radio pulsations, a quarter are detected only in  high-energy pulsations and a quarter only in radio pulsations, with no particular distribution on the period-period derivative plane. 
There are only two other known extragalactic pulsars in supernova remnants. Both are in Pulsar Wind Nebulae of the Large Magellanic Cloud. PSR J0537$-$6910, a fast (16\,ms) pulsar, is only\footnote{Unpulsed gamma-rays from the PWN have been detected \citep{Abramowski2012}.} detected in X-rays \citep{Marshall1998}, just like \thepulsar{}. PSR B0540$-$6919 is detected in X-rays, radio and gamma-rays and is considered an analogue of the Crab pulsar \citep{Seward1984,Manchester1993, Marshall2016}. It is a giant pulse emitter \citep{Johnston2003}. It has a very low radio efficiency of $\eta_{\text{radio}}= 2.9 \times 10^{-9}$, of the same order as the upper limit on the radio efficiency of \thepulsar{}. There is only one other extragalactic young\footnote{All other published non-magnetar extragalactic pulsars are over 200\,kyr in characteristic age \citep{ATNF}.} non-magnetar pulsar  known: J0534$-$6703 \citep{Manchester2006}. It is a long period ($\simeq$1.8\,s) radio pulsar with no SNR or PWN association and no high-energy emission known. There are no confirmed Central Compact Objects (CCO) in the Magellanic Clouds \citep[e.g.][]{Long2020}. CCOs  are  soft thermal X-ray sources  near the centre of a supernova remnant, with no hard X-ray emission, radio emission, or  PWN environment \citep{DeLuca2017}. 

\section{Conclusion}
We conclude from our unprecedentedly deep radio search that the X-ray pulsar \thepulsarnopsr{} is either similar to a handful of pulsars in Pulsar Wind Nebulae with very low radio efficiency or its radio beam does not cross our line of sight. The latter hypothesis is supported by the non-detection of thermal X-rays from the polar cap of \thepulsar{} in \cite{Maitra2021}, and the non-detection of giant pulses in this study. Only about half of young pulsars with SNR associations have been detected in both high-energy and radio pulsations. The upper limit set on \thepulsar{}'s radio flux density in this work is nearly 7 times deeper than previous searches. The MeerKAT full array, or the future Square Kilometer Array, could carry out a deeper search on this source.  The MeerKAT full array could improve the search sensitivity by a factor of $\simeq1.5$ compared to this work with 20 additional dishes. If \thepulsar{} was not detected in a deeper search with the full array, this would not significantly strengthen the case for a non-alignment, as its radio efficiency upper limit would still be higher than 7 pulsars. There are only about 20 radio pulsar-PWN systems in the Milky Way. If it was detected, \thepulsar{} would be the second extragalactic radio pulsar-PWN system (after PSR B0540$-$6919 in the Large Magellanic Cloud, \cite{Manchester1993}), and the first in the SMC.

\section*{Acknowledgements}
The MeerKAT telescope is operated by the South African Radio Astronomy Observatory, which is a facility of the National Research Foundation, an agency of the Department of Science and Innovation. SARAO acknowledges the ongoing advice and calibration of GPS systems by the National Metrology Institute of South Africa (NMISA) and the time space reference systems department of the Paris Observatory.

TRAPUM observations used the FBFUSE and APSUSE computing clusters for data acquisition, storage and analysis. These clusters were funded and installed by the Max-Planck-Institut für Radioastronomie and the Max-PlanckGesellschaft.

EC acknowledges funding from the United Kingdom's Research and Innovation Science and Technology Facilities Council (STFC) Doctoral Training Partnership, project reference 2487536. For the purpose of open access, the author has applied a Creative Commons Attribution (CC BY) licence to any Author Accepted Manuscript version arising. 

EB, MK, PVP and VVK acknowledge continuing support from the Max Planck society. 

RPB acknowledges support from the ERC under the European Union's Horizon 2020 research and innovation programme (grant agreement No. 715051; Spiders).

We thank Fabian Jankowski and James Turner (Pulsar and Time-Domain Astrophysics group, Jodrell Bank Centre for Astrophysics, University of Manchester) for their help with the primary beam model and the sky temperature estimation respectively. 

This paper has made use of the ATNF pulsar catalogue version 1.67.

We thank the anonymous referee for their helpful and constructive comments which improved this manuscript.

\section*{Data Availability}
The data underlying this article will be shared upon reasonable request to the TRAPUM collaboration.

\bibliographystyle{mnras}
\bibliography{main}

\bsp	
\label{lastpage}
\end{document}